\documentclass[a4paper]{raa} 

\usepackage{graphicx,times}             
\usepackage{natbib}
\usepackage{amssymb,amsmath}
\bibpunct{(}{)}{;}{a}{}{,}

\usepackage[pagebackref=true]{hyperref}

\usepackage{makecell}
\usepackage{multirow}
\usepackage{ulem}
\usepackage{xcolor}
\usepackage{float}

\newcommand{\kms}{km s$^{-1}$}
\newcommand{\dego}{$^\circ$}
\newcommand{\msun}{M$_\odot$}

\begin{document}

  \title{Contributions of rotation, expansion and line broadening to the morphology and kinematics of the inner CSE of oxygen-rich AGB star \mbox{R Hya}
}

   \volnopage{Vol.0 (20xx) No.0, 000--000}      
   \setcounter{page}{1}          

   \author{Pham Tuyet Nhung
      \inst{}
   \and Do Thi Hoai
      \inst{}
   \and Pierre Darriulat
      \inst{}
   \and Pham Ngoc Diep
      \inst{}
   \and Nguyen Bich Ngoc
      \inst{}
   \and Tran Thi Thai
      \inst{}
   \and Pham Tuan-Anh
      \inst{}
   }

   \institute{Department of Astrophysics, Vietnam National Space Center, Vietnam Academy of Science and Technology, 18, Hoang Quoc Viet, Nghia Do, Cau Giay, Ha Noi, Vietnam; \\
{\it pttnhung@vnsc.org.vn, dthoai@vnsc.org.vn}\\
\vs\no
   {\small Received 20xx month day; accepted 20xx month day}}

\abstract{ We use archival ALMA observations of the CO(2-1) and SiO(5-4) molecular line emissions of AGB star R Hya to illustrate the relative contributions of rotation, expansion and line broadening to the morphology and kinematics of the circumstellar envelope (CSE) within some $\sim$70 au ($\sim$0.5 arcsec) from the centre of the star. We give evidence for rotation and important line broadening to dominate the inner region, within $\sim$14 au ($\sim$100 mas) from the centre of the star. The former is about an axis that projects a few degrees west of north and has a projected rotation velocity of a few \kms. The latter occurs within some 7-14 au (50-100 mas) from the centre of the star, the line width reaching two to three times its value outside this region. We suggest that it is caused by shocks induced by stellar pulsations and convective cell granulation. We show the importance of properly accounting for the observed line broadening when discussing rotation and evaluating the radial dependence of the rotation velocity.
\keywords{stars: AGB and post-AGB --- circumstellar matter --- stars: individual: R Hya --- radio lines: stars}
}

   \authorrunning{P.T. Nhung et al.}       
   \titlerunning{Inner CSE of AGB star R Hya}  

   \maketitle

%
%
\section{Introduction\label{intro}}        

The inner layers of the circumstellar envelope (CSE) of oxygen-rich Asymptotic Giant Branch (AGB) stars have often been shown to display both rotation and turbulence. The availability of high angular resolution observations using the Very Large Telescope (VLT) in the visible and infrared and the Atacama Large Millimeter/submillimeter Array (ALMA) at millimetre wavelengths have made such observations possible. The latter observe the movement of the gas in the inner layers of the CSE from the detection of the emission of its molecular lines. Rotation is identified as displaying a characteristic position-velocity (PV) pattern and turbulence as causing a significant broadening of the line widths.

Rotation is often interpreted as suggesting the presence of a planetary or stellar companion and line broadening as being the result of shocks induced by stellar pulsations and convective cell granulation \citep{Hofner2019}. Examples of rotation are R Dor \citep{Vlemmings2018, Homan2018, Hoai2020, Nhung2021} and L$_2$ Pup \citep{Kervella2016, Homan2017, Hoai2022} and examples of line broadening are again R Dor and o Cet \citep{Nhung2022}. The interplay between rotation and line broadening complicates the analysis of these observations in a way that can easily be underestimated. The aim of the present article is to illustrate this point using as an example the oxygen-rich AGB star R Hya.

R Hya is a Mira-type pulsator of spectral type \mbox{M6-9} \citep{Wenger2000} with a period of 385 days \citep{Zijlstra2002}. Hipparcos estimates a distance of $124\substack{+12 \\ -10}$\,pc \citep{vanLeeuwen2007} and Gaia EDR3 of $148\substack{+11 \\ -10}$\,pc \citep{Gaia2021}. Stellar diameters are $\sim$30 mas in the near-infrared \citep{Haniff1995} and $\sim$23 mas in the visible \citep{Richichi2005}. An effective temperature of ~2100 K has been measured by \citet{DeBeck2010}. 

Recently, the circumstellar environment of the star was observed with ALMA as part of the ATOMIUM large programme \citep{Homan2021}. The target was observed with three different antenna configurations, covering radial distances from $\sim$5 to $\sim$1500 au, allowing for an in-depth morphological study of the stellar wind using maps of continuum emission and of CO(2-1) and SiO(5-4) molecular line emissions. These observations reveal a very complex mass loss history but, in the present article, we limit our exploration to the inner layers confined within $\sim$500 mas from the star. In this region, CO emission suggests the presence of an equatorial density enhancement and SiO emission reveals that the gas reaches Doppler velocities of up to $\pm$18 \kms, interpreted by \citet{Homan2021} as the nearly Keplerian rotation of a ring-like feature, having a radius of $\sim$6 au, and hosting a companion with a mass of some 0.65 \msun.

\section{Observations and data reduction\label{sec2}}

\begin{figure*}
  \centering
  \includegraphics[width=4.6cm,trim=0cm .5cm 1cm 1.7cm,clip]{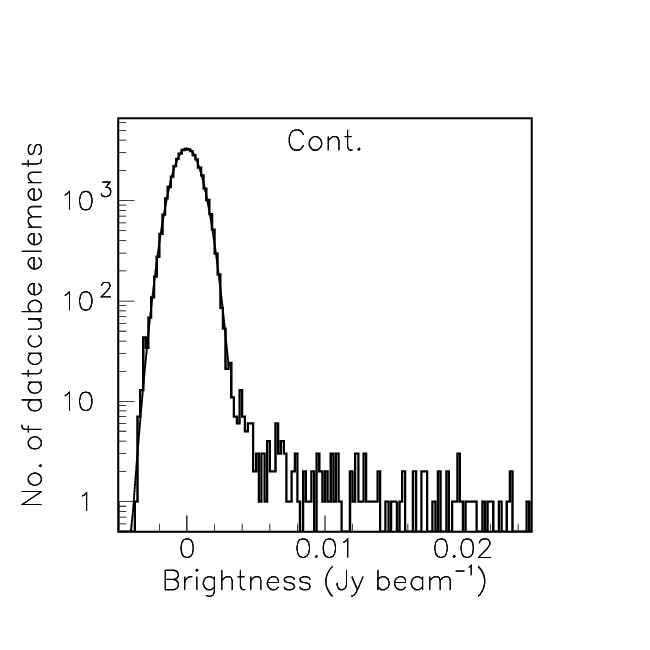}
  \includegraphics[width=4.6cm,trim=0cm .5cm 1cm 1.7cm,clip]{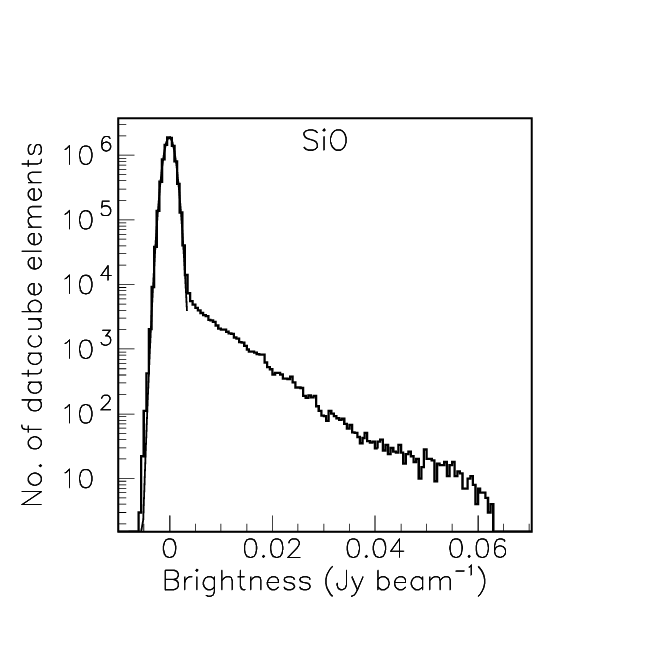}
  \includegraphics[width=4.6cm,trim=0cm .5cm 1cm 1.7cm,clip]{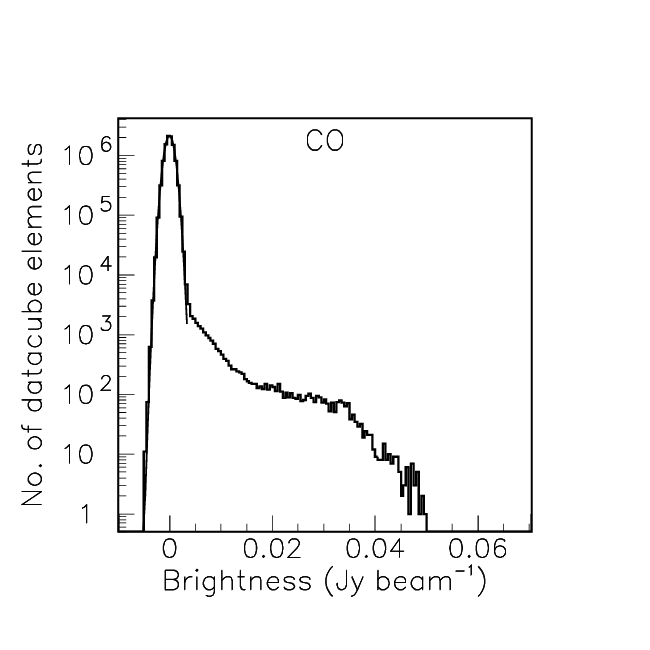}
  \caption{Brightness distributions measured in $|x,y|<0.6$ arcsec for continuum emission (left), and in $R<1.5$ arcsec and $|V_z|<32$ \kms\, for SiO(5-4) (centre) and CO(2-1) (right) emissions.}
 \label{fig1}
\end{figure*}

We make exclusive use of the extended antenna configuration of the ATOMIUM observations \citep[ADS/JAO.ALMA\#2018.1.00659.L,][]{Homan2021}, made on 11 July 2019 and using 44 antennas with baselines covering between 0.11 and 12.6 km, corresponding to a maximum recoverable scale\footnote{Maximum Recoverable Scale (MRS): the largest angular structure to which a given array is sensitive. It is in principle determined by the shortest baseline of the array configuration. } of 0.6 arcsec. We refer the reader to \citet{Homan2021} and \citet{Gottlieb2022} for details. Images of the SiO(5-4) emission (frequency of 217.1049 GHz, spectral window 27) and of the CO(2-1) emission (frequency of 230.5380 GHz, spectral window 31) have been produced, without preliminary continuum subtraction, using GILDAS\footnote{https://www.iram.fr/IRAMFR/GILDAS/}/IMAGER\footnote{https://imager.oasu.u-bordeaux.fr} with robust weighting (weight=1), resulting in beams of 64 mas $\times$ 30 mas (FWHM), PA=66\dego, for the SiO line and 58 mas $\times$ 28 mas, PA=65\dego, for the CO line. We note that the elongation of the beam is 20\% larger using GILDAS than using CASA\footnote{https://casa.nrao.edu/} for imaging, but we do not know what is causing the difference. Figure \ref{fig1} displays brightness distributions for continuum emission in a square $|x,y|<0.6$ arcsec, with $x$ pointing east and $y$ pointing north, and for molecular line emissions in a circle $R=(x^2+y^2)^\frac{1}{2}<1.5$ arcsec. They correspond to a noise level ($\sigma$) of 0.98 mJy beam$^{-1}$ for continuum and the SiO line and 0.90 mJy beam$^{-1}$ for the CO line. Doppler velocities, $V_z$, are referred to a systemic velocity of $-$10.1 \kms. 

\section{Morphology and kinematics of the inner CSE \label{sec3}}
\subsection{Overview\label{sec3:ss1}}
\begin{figure*}
  \centering
  \includegraphics[width=4.6cm,trim=0cm .5cm 1cm 0cm,clip]{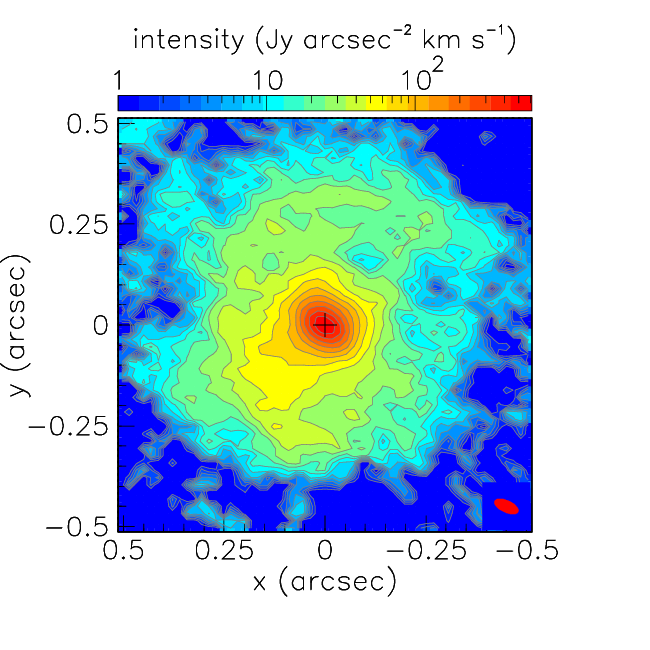}
  \includegraphics[width=4.6cm,trim=0cm .5cm 1cm 0cm,clip]{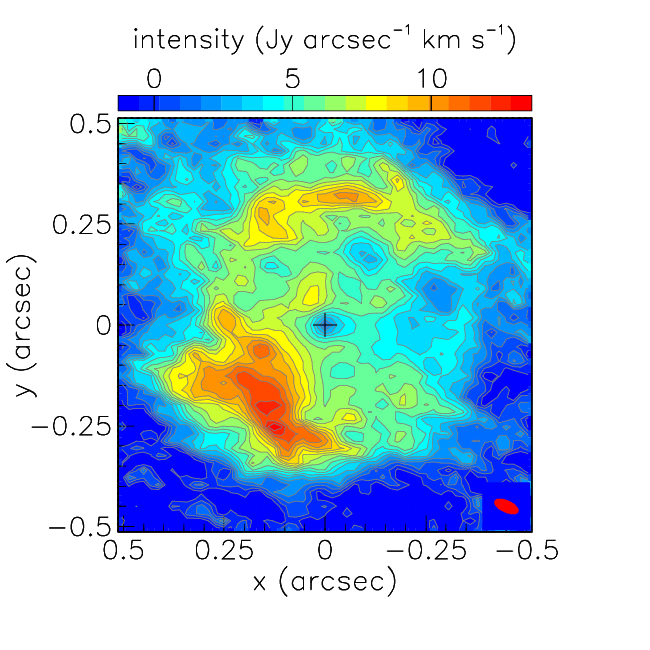}
  \includegraphics[width=4.6cm,trim=0cm .5cm 1cm 0cm,clip]{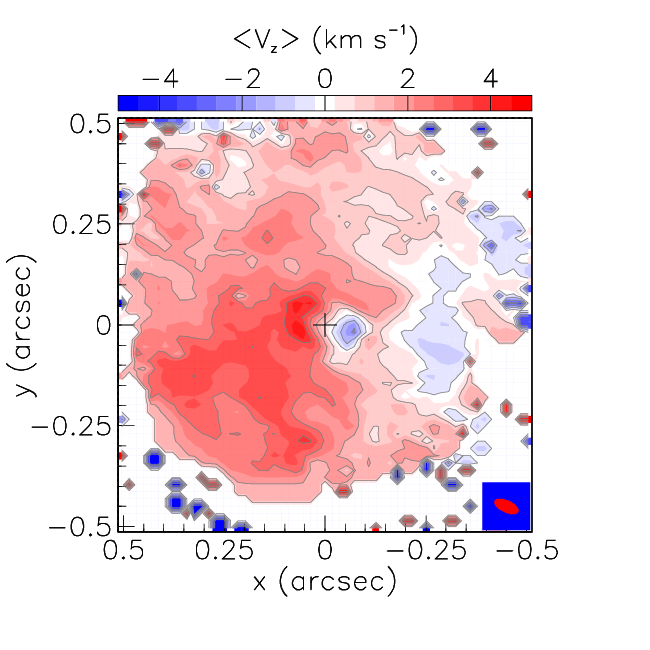}
  \includegraphics[width=4.6cm,trim=0cm .5cm 1cm 1.7cm,clip]{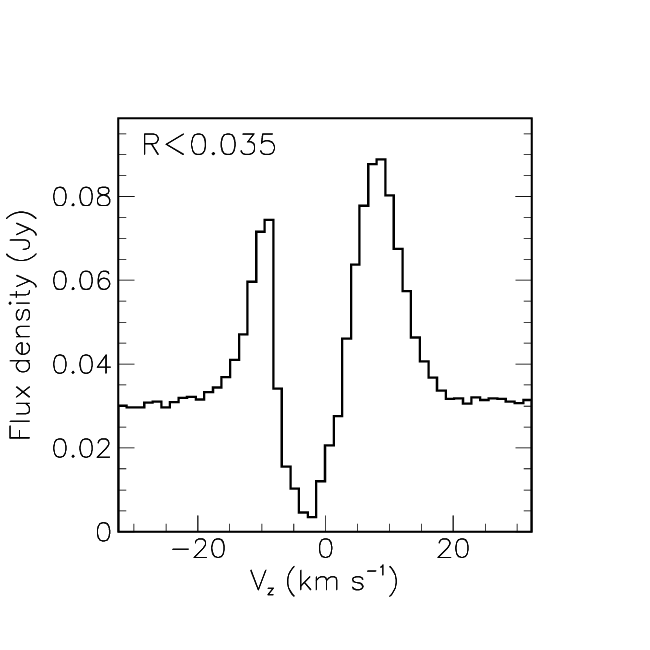}
  \includegraphics[width=4.6cm,trim=0cm .5cm 1cm 1.7cm,clip]{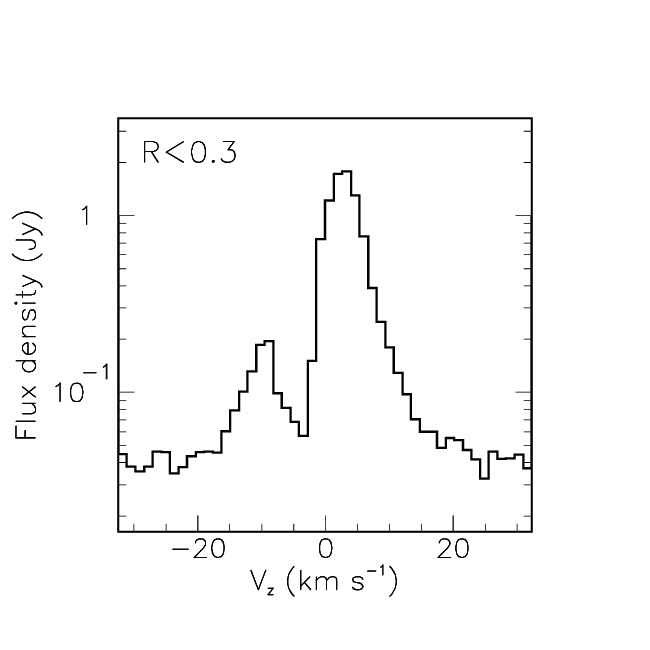}
  \includegraphics[width=4.6cm,trim=0cm .5cm 1cm 1.7cm,clip]{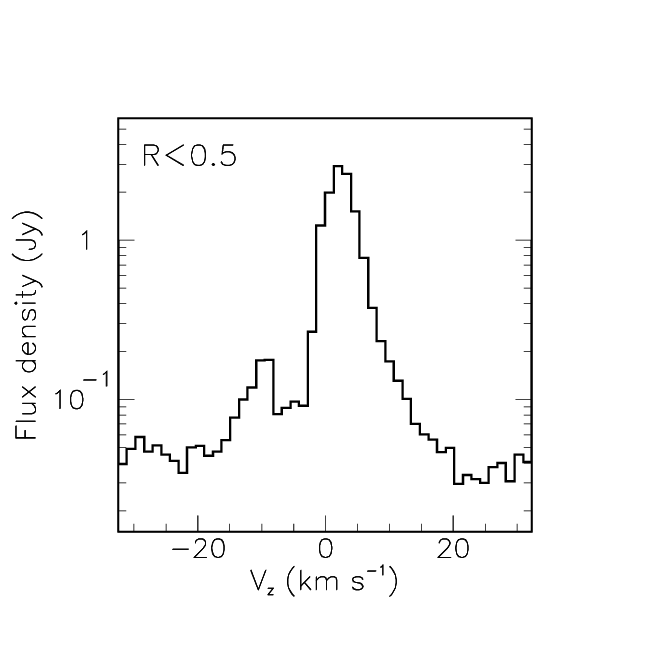}
  \caption{SiO(5-4) emission of the CSE of R Hya. The upper row shows maps of the intensity (left and centre), integrated over $|V_z|<20$ \kms\ (left) and $|V_z|<8$ \kms\ (centre) and of the mean Doppler velocity (right), averaged over $|V_z|<20$ \kms. The central map displays the intensity multiplied by $R$, which would produce a uniform brightness distribution for an optically thin density decreasing in inverse proportion to the distance squared. The lower row shows Doppler velocity spectra integrated over circles of different radii: 35 mas (left), 0.3 arcsec (centre) and 0.5 arcsec (right).}
 \label{fig2}
\end{figure*}

Figure \ref{fig2} displays maps of the intensity (moment 0) and mean Doppler velocity (moment 1) of the emission of the SiO(5-4) line, together with Doppler velocity spectra.

The intensity maps display a clear enhancement of emission in the south-eastern quadrant, corresponding to red-shifted velocities of a few \kms. In the three other quadrants, the intensity is enhanced in the northern and north-western directions but depressed in the western direction and the Doppler velocity, averaged over the three quadrants using intensity as weight, approximately cancels. The lack of emission at some 300 mas from the star at position angles between $\sim$270\dego\ and $\sim$300\dego\ (measured counter-clockwise from north) was noted by \citet{Homan2021} and interpreted as a dense clump of cool gas or as an exceptionally low-density region, not excluding, however, that it might be caused by photo-dissociation by a hidden white dwarf companion. Close to the star, in a region confined within $R<$ $\sim$100 mas, the moment 1 map displays a pattern typical of rotation (see Section 3.2), with an axis projecting on the plane of the sky at a position angle of a few degrees west of north, consistent with the estimate of $-$20\dego\ obtained by \citet{Homan2021}. 

The Doppler velocity spectra are integrated over circles of different radii, 35 mas, 300 mas and 500mas. The contribution of the continuum is at the level of 0.03-0.04 Jy. High $V_z$ wings are seen on all three spectra, extending to $\pm$$\sim$20 \kms. Very strong absorption is seen in the blue-shifted hemisphere, nearly total in the smaller circle, revealing the high opacity of the SiO line, a well-known result of the conjunction between a large density and a very high Einstein coefficient ($\sim$5.2$\times$10$^{-4}$ Hz \citep{Muller2005}). The absorption peaks cover an interval of Doppler velocity possibly suggesting a progressive radial acceleration to a terminal velocity of some 7 \kms.

The remaining of this section discusses separately the inner region, $R<100$ mas, in sub-section \ref{sec3:ss2} and the region $R>100$ mas in sub-section \ref{sec3:ss3}.

\subsection{The inner region: rotation and line broadening\label{sec3:ss2}}

\begin{figure*}
  \centering
  \includegraphics[height=5.cm,trim=0cm .5cm -1.cm 1.7cm,clip]{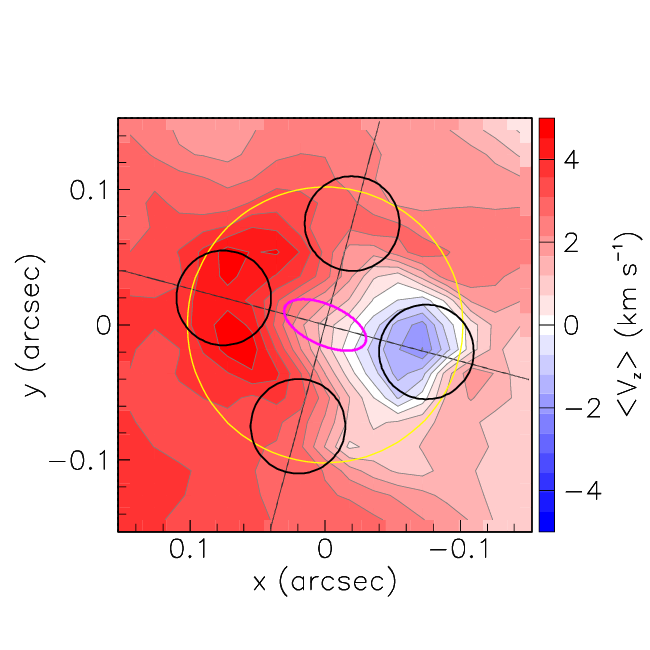}
  \includegraphics[height=5.cm,trim=0cm .5cm 0cm 1.7cm,clip]{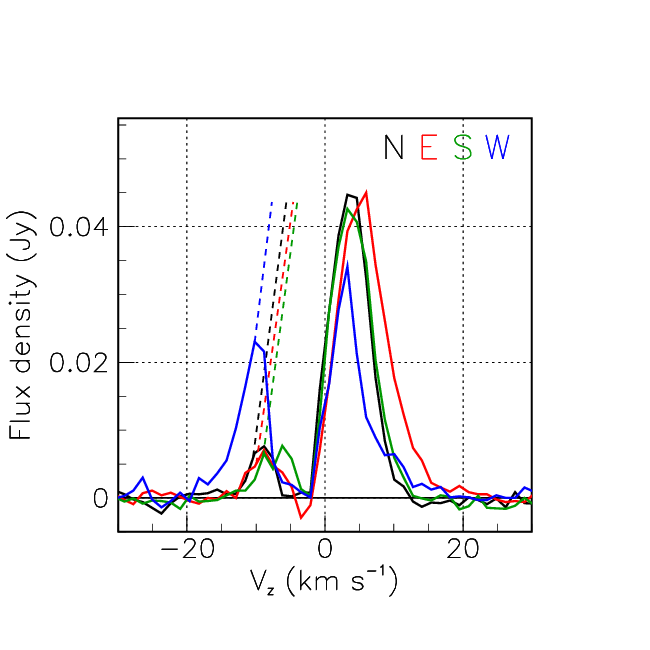}
  \caption{Left: moment 1 map of the SiO(5-4) emission in the inner region. Black circles indicate the regions over which the Doppler velocity spectra displayed in the right panel have been integrated. The yellow circle delimitates the $R$$<$100 mas region discussed in Section 3.2. The magenta ellipse in the centre of the map shows the beam. Right: Doppler velocity spectra integrated over the regions indicated in the left panel. N, E, S and W refer to position angles of $-$20\dego, 70\dego, 160\dego and 250\dego, respectively. Dotted lines are simply extrapolating the blue wings with the slopes observed on the red-shifted side and are only meant to guide the eye.}
 \label{fig3}
\end{figure*}

Figure \ref{fig3} displays a zoom on the central part of the moment 1 map of the SiO(5-4) emission. It shows a clear rotation pattern about an axis projecting a few degrees west of north and a projected rotation velocity of a few \kms. Assuming the position angle of the projected rotation axis to be $-$20\dego\ as estimated by \citet{Homan2021}, we have selected four regions of a size comparable to the beam, at a mean projected distance of $\sim$80 mas from the centre of the star, for which we display, in the right panel, Doppler velocity spectra. They show broad line profiles, covering up to $\sim$30 \kms\ at the base, respectively red- and blue-shifted in the regions at mean position angles of 70\dego\ and 250\dego. The spectra observed on the projection of the rotation axis are centred in between these two. We evaluate the shift between the blue-shifted and red-shifted spectra as 5$\pm$1 \kms, meaning a projected rotation velocity of 2-3 \kms.

\begin{figure*}
  \centering
  \includegraphics[width=4.6cm,trim=0cm .5cm 1cm 0cm,clip]{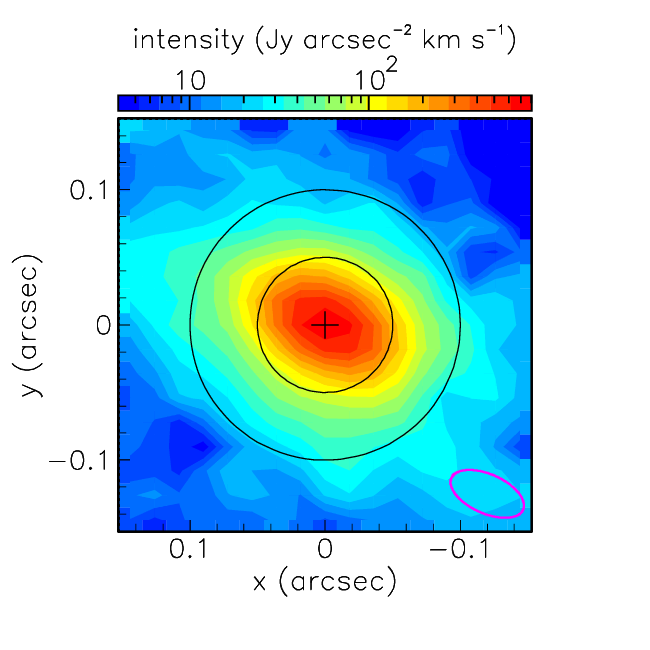}
  \includegraphics[width=4.6cm,trim=0cm .5cm 1cm 0cm,clip]{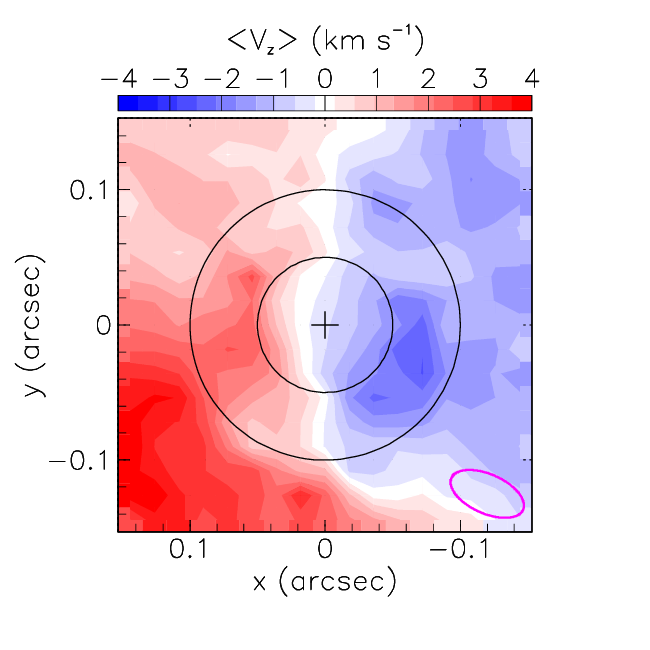}
  \includegraphics[width=4.6cm,trim=0cm .5cm 1cm 0cm,clip]{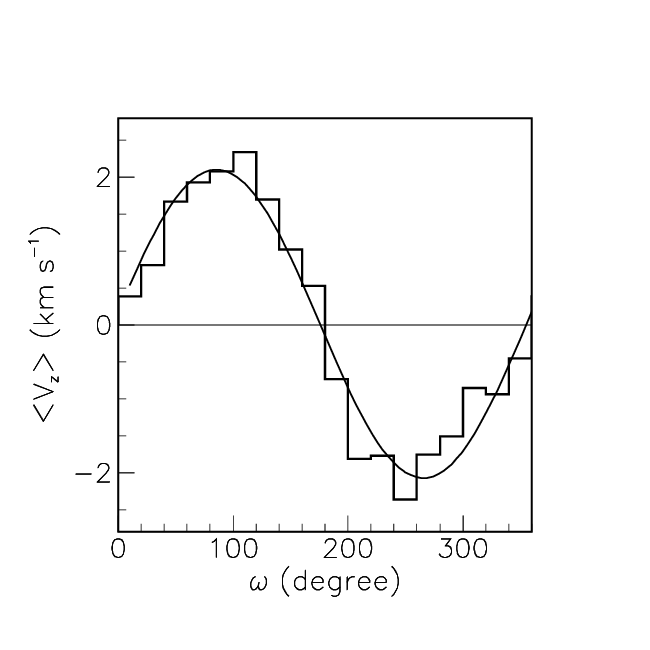}
  \caption{CO(2-1) emission in the inner 100 mas region. Left: moment 0 map integrated over $|V_z|<20$ \kms. Centre: moment 1 map averaged over the same interval. Right: sine wave fit to the dependence on position angle $\omega$ of the intensity-weighted mean Doppler velocity measured in the ring $50<R<100$ mas.}
 \label{fig4}
\end{figure*}

As a further evidence for the dominance of rotation in the inner 100 mas, we show in Figure \ref{fig4} the intensity and moment 1 maps of the CO(2-1) emission, which is much less opaque than the SiO(5-4) emission and allows for reliable evaluations of the mean Doppler velocity in each pixel. The right panel of the figure displays its dependence on position angle $\omega$ averaged over the ring $50<R<100$ mas. A sine wave fit gives $\langle V_z \rangle=2.1\sin(\omega+4^\circ)$ \kms. This suggests a position angle of the rotation axis of $-$4\dego, with an uncertainty of at least $\pm$10\dego. This is closer to north by 16\dego\ than the value of $-20$\dego\ evaluated by  \citet{Homan2021}, a difference that is acceptable within errors.

\begin{figure*}
  \centering
  \includegraphics[width=4.6cm,trim=0cm .5cm 1cm 0cm,clip]{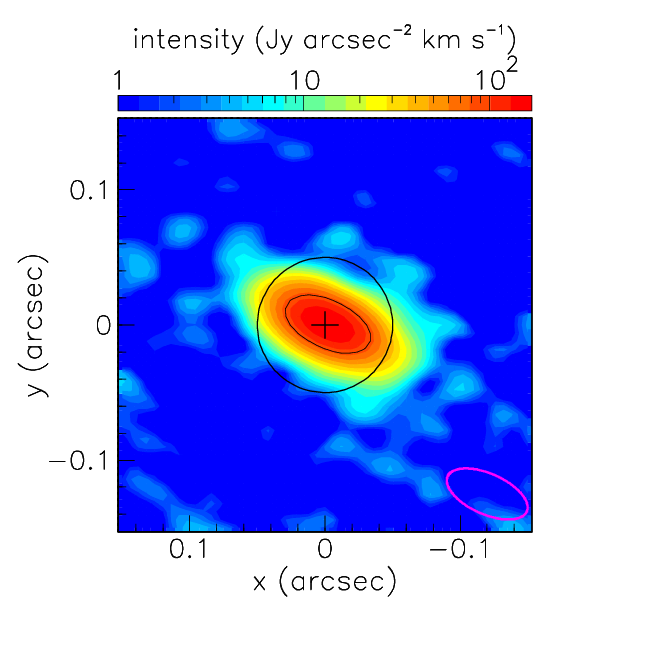}
  \includegraphics[width=4.6cm,trim=0cm .5cm 1cm 0cm,clip]{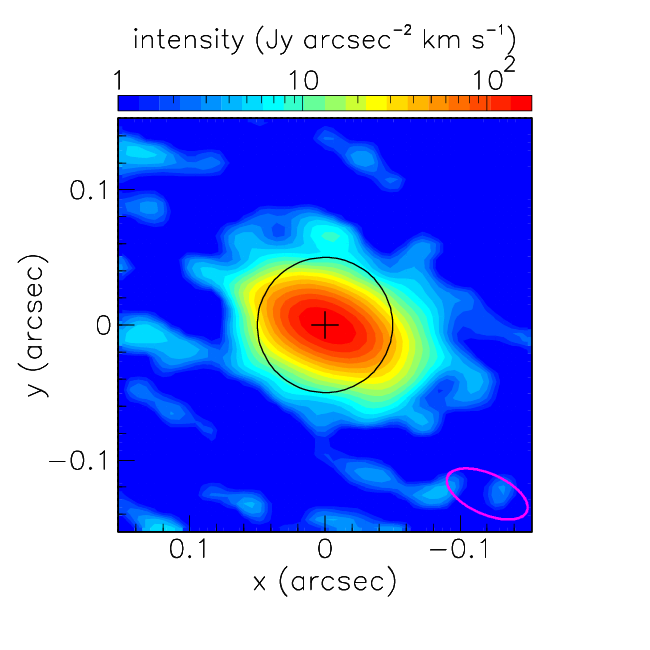}
  \includegraphics[width=4.6cm,trim=0cm .5cm 1cm 0cm,clip]{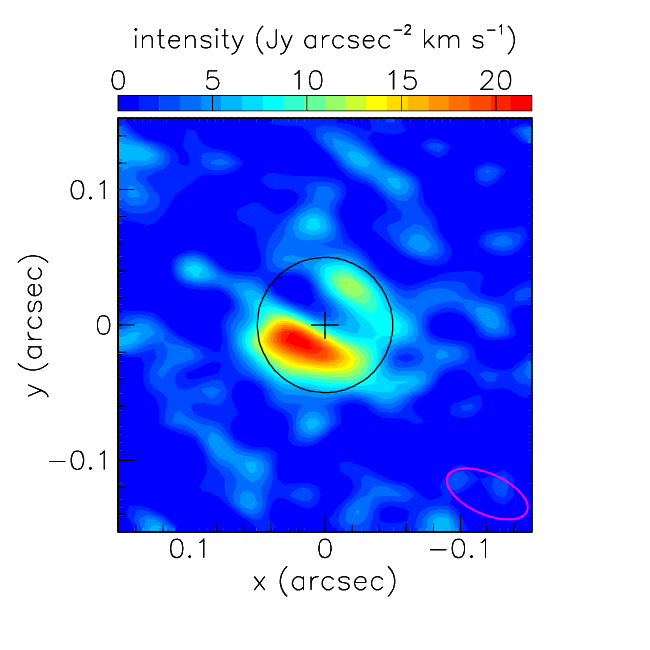}
  \caption{Intensity maps of continuum emission ($23<|V_z|<28$ \kms, left), of high $|V_z|$ SiO emission ($15<|V_z|<20$ \kms, centre) and of their difference (right). The black contour shown on the continuum intensity distribution is at half maximum emission and is comparable in size to the half-maximum beam contour shown in the lower right corner, indicating that the observed ellipticity of the continuum map is the result of the convolution with the beam. In each panel, the cross locates the centre of the star; the black circle is centred on the star and has a radius of 50 mas; and the magenta ellipse shows the beam.}
 \label{fig5}
\end{figure*}

Evidence for the confinement of the line broadening region to short distances from the centre of the star is illustrated in Figure \ref{fig5}. It displays intensity maps of the high $V_z$ wings, defined as $15<|V_z|<2$0 \kms, of the continuum, defined as $23<|V_z|<28$ \kms, and of their difference. The latter is seen to be well contained in a circle of radius $\sim$50 mas centred on the star. The strong opacity of the SiO line implies that subtracting the contribution of the continuum emission cannot be simply obtained by subtracting the respective intensities but would require a radiative transfer simulation, which is beyond the scope of the present article. As a result, the difference between the intensities of the two maps takes negative values over and near the stellar disc. However, on the limbs of the region of high $|V_z|$ SiO emission, this argument does not apply and the subtracted map provides a reliable estimate of its edges, making the evidence for confinement reliable and strong. Note that the subtracted intensity is an order of magnitude smaller than the terms of the subtraction, implying important uncertainties. In particular, the apparent non uniformity of the subtracted intensity over the stellar disc, displaying a clear enhancement in the south-eastern hemisphere, would require observations of higher sensitivity and better angular resolution to be considered as a reliable result.  

\begin{figure*}
  \centering
  \includegraphics[width=13.5cm,trim=0cm .5cm 1cm 0cm,clip]{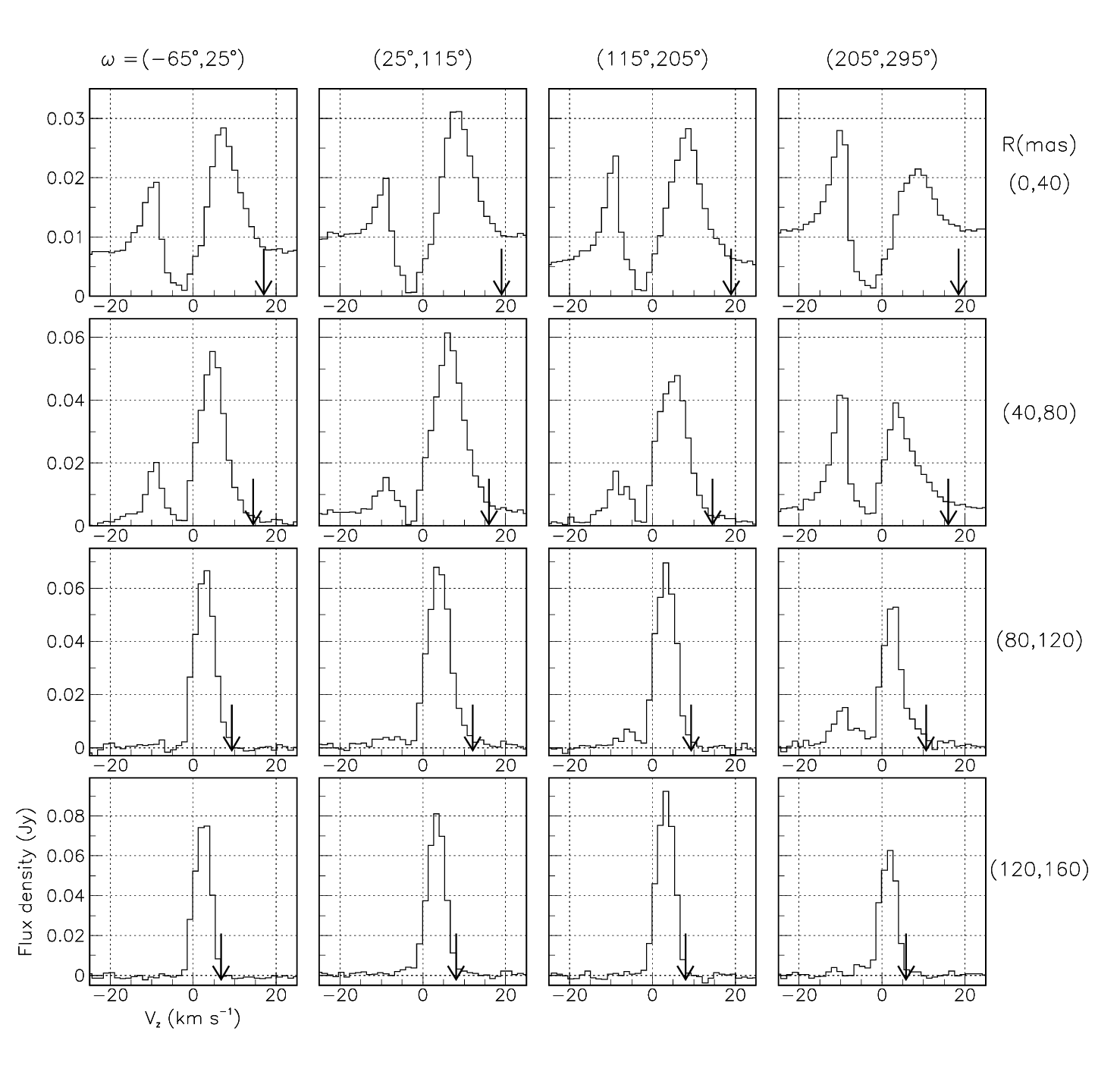}
  \caption{Doppler velocity spectra of the SiO emission measured in rings and quadrants defined in the text. The rows, from up down, correspond to the successive rings from inside outwards. The columns, from left to right, correspond to the successive quadrants, starting from the quadrant centred at $-$20\dego\ and moving counter-clockwise. The arrows show the end points of each individual spectrum.}
 \label{fig6}
\end{figure*}

At this point, we have obtained evidence for the joint presence, in the $R$$<$100 mas inner region of the CSE, of three features that affect the line profiles: opacity, particularly strong on the SiO line, seen as a deep absorption peak on the Doppler velocity spectra illustrated in Figures \ref{fig2}, \ref{fig3}, \ref{fig6} and discussed in the third paragraph of Section \ref{sec3:ss1}; rotation, with a projected velocity of $\sim$2 \kms\ and an axis projecting $\sim$4\dego$\pm$10\dego west of north, seen on both the SiO and CO lines; line broadening, seen as the presence of high $|V_z|$ wings on the line profiles, confined within $<$50 mas from the centre of the star. As they may influence each other, we need to make sure that their effects can be reliably disentangled. Experience in having addressed this issue in other AGB stars is a precious asset, in particular because the relative importance of each of the three features is different in different stars. If the angular resolution were much smaller than the region being explored, say at the level of 1 mas compared with 100 mas, and if the line profile were not much distorted by the presence of absorption, the problem would be easily solved: at each ($x$,$y$) point of the intensity map, the width of the line profile would provide a direct measurement of the line broadening and the dependence of its mean Doppler velocity on position angle a direct measurement of the projected rotation velocity and of the orientation of the projected rotation axis. In practice, however, the angular resolution has a mean $\sigma$ of 20 mas and the absorption of the SiO line is very strong over the range of Doppler velocities across which the wind is accelerated, $\sim-$3$\pm$3 \kms. The dependence on $r$, the space distance to the centre of the star, of the rotation velocity and of the line broadening is poorly known. There is no reason to assume that the former would be Keplerian: such a regime applies when the acceleration of gravity is exactly balanced by the centrifugal acceleration, as is the case for a planet or, to a lesser extent, for gas orbiting around a protostar, but is not expected to apply in the inner atmosphere of an AGB star, where pulsations, shocks and interactions of the stellar UV radiation with newly formed dust grains play an important role. Concerning line broadening, what is known with certainty is that it is confined within short distances from the star, but its precise dependence on $r$ is not known. To cope with these difficulties, it is important to study the dependence of the observed line profile on its location on the intensity map, remarking in particular that the effect of rotation cancels on the projection of the rotation axis and is maximal in the perpendicular direction. This is done in Figure \ref{fig6}, which displays velocity spectra in four quadrants of four rings. 
The rings cover successively radial intervals [0,40], [40,80], [80,120] and [120,160] mas and the quadrants cover position angle intervals [$-$65\dego,25\dego], [25\dego,115\dego], [115\dego,205\dego] and [205\dego,295\dego]. In each spectrum we evaluate the Doppler velocity of the end point of the spectrum (above continuum) on the red side where absorption is unimportant. The results, averaged in each ring, are, from inside outwards, 18.5$\pm$0.9 \kms, 15.0$\pm$0.8 \kms, 10.5$\pm$1.2 \kms\ and 7.1$\pm$0.7 \kms. The values quoted after the $\pm$ sign are the rms deviations with respect to the mean for the four quadrants of a same ring. This result implies a broadening of $\sim$25$\pm$2 \kms\ at the base of the line over a radial distance of $\sim$120 mas (from 20 to 140 mas), half of it being reached at $R$$\sim$80 mas.

Accounting for the beam size, we can safely retain from the above considerations that significant line broadening occurs within some 50-100 mas from the centre of the star, the line width reaching two to three times its value outside this region.  

\subsection{The outer region: radial expansion\label{sec3:ss3}}

\begin{figure*}
  \centering
  \includegraphics[width=4.6cm,trim=0cm .5cm 1cm 0cm,clip]{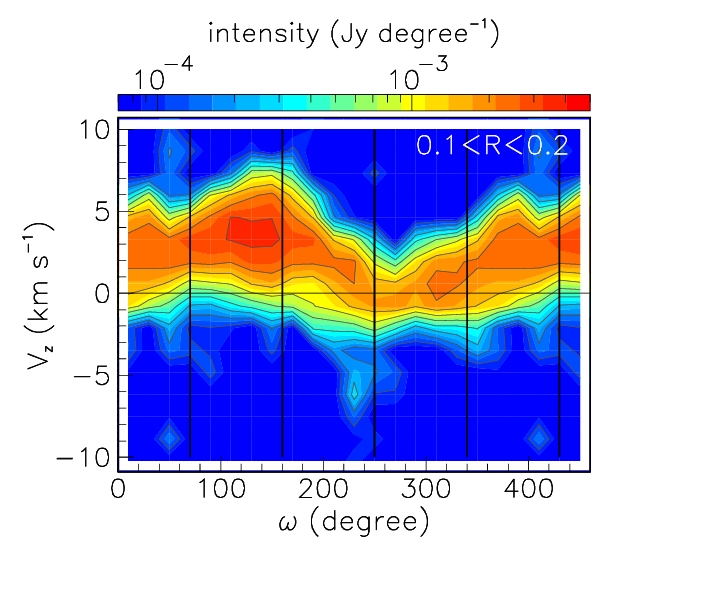}
  \includegraphics[width=4.6cm,trim=0cm .5cm 1cm 0cm,clip]{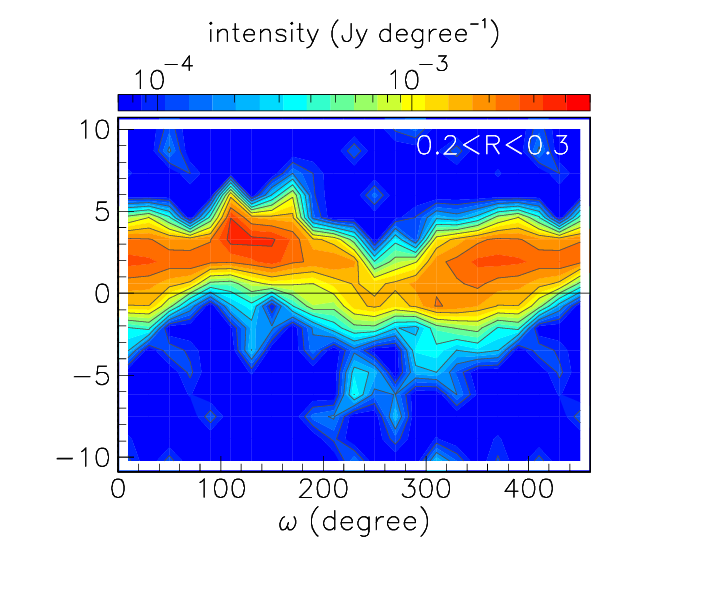}
  \includegraphics[width=4.6cm,trim=0cm .5cm 1cm 0cm,clip]{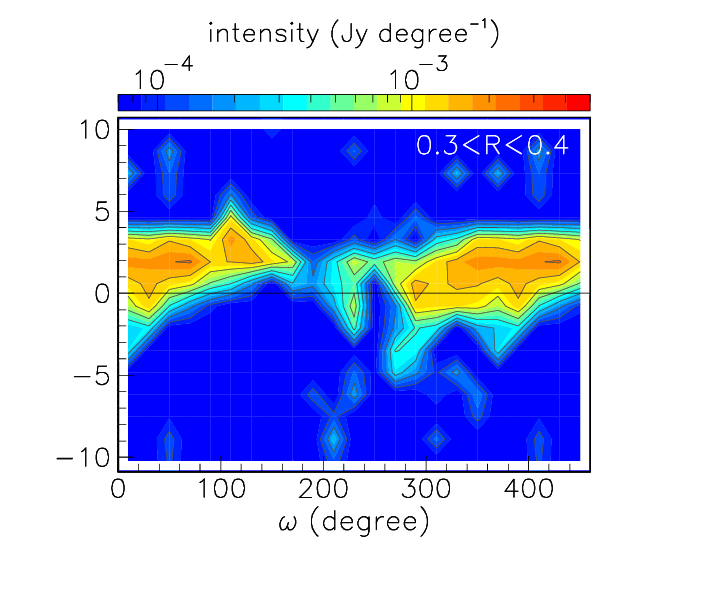}
  \caption{PV maps, $V_z$ vs $\omega$, of the SiO(5-4) emission in three successive rings as indicated in the upper-right corner of each panel. In the left panel, vertical lines show position angles of 70\dego\ modulo 90\dego.}
 \label{fig7}
\end{figure*}

Figure \ref{fig7} displays PV maps of the SiO(5-4) emission, Doppler velocity $V_z$ vs position angle $\omega$, integrated in three successive rings: $0.1<R<0.2$ arcsec, $0.2<R<0.3$ arcsec and $0.3<R<0.4$ arcsec. It shows a pattern typical of an equatorial enhancement expanding radially at a velocity of a few \kms\ and inclined with respect to the plane of the sky with the red-shifted part projecting at position angles centred on $\sim$120\dego\ and the blue-shifted part projecting back-to-back at projection angles centred on $\sim$300\dego. It is already clear in the smaller ring, $0.1<R<0.2$ arcsec, that the rotation regime at stake in the inner region, $R<0.1$ arcsec, does no longer apply: it would imply $V_z$ to be maximal at 70\dego, to cancel at 160\dego, to be minimal at 250\dego\ and to cancel again at 340\dego, which is not at all the case. The map of the larger ring shows clearly the hole of emission noted by \citet{Homan2021} but also, more generally, a broad depression covering the whole south-western quadrant. Understanding what is causing it would require additional observations of better sensitivity and of other molecular lines.

\section{Summary and discussion}
The picture of the inner CSE region of R Hya presented in the previous section can be qualitatively summarised as follows: a central region, confined to some $\sim$7 stellar radii ($\sim$100 mas or 14 au) from the centre of the star, is dominated by rotation and by important line broadening. The former is about an axis that projects a few degrees west of north and has a projected rotation velocity of a few \kms. The latter occurs within some 7-14 au from the centre of the star, the line width reaching two to three times its value outside this region. Such broadening is too large for standard sources of line broadening (essentially thermal broadening) and is probably the result of shocks induced by stellar pulsations and convective cell granulation as suggested by current state-of-the art hydrodynamical models \citep[e.g.] []{Hofner2019}. However, to give a precise meaning to this statement would require a detailed analysis accounting for the scale of the velocity gradient in both space and time, which is well beyond the scope of the present study. Beyond the central region, at radial distances between 100 and 500 mas, the wind is progressively accelerated radially, with a red-shifted enhancement in the south-eastern quadrant. The transition between the two regimes is not abrupt and some rotation may persist beyond $R$$\sim$14 au ($\sim$100 mas) while some radial expansion may be present in the central region. The terminal velocity reached within this region is $\sim$7 \kms, which does not exclude, however, that higher values be obtained at larger distances from the star.   

\begin{figure*}
  \centering
  \includegraphics[width=4.8cm,trim=0.5cm .5cm 1.7cm 0cm,clip]{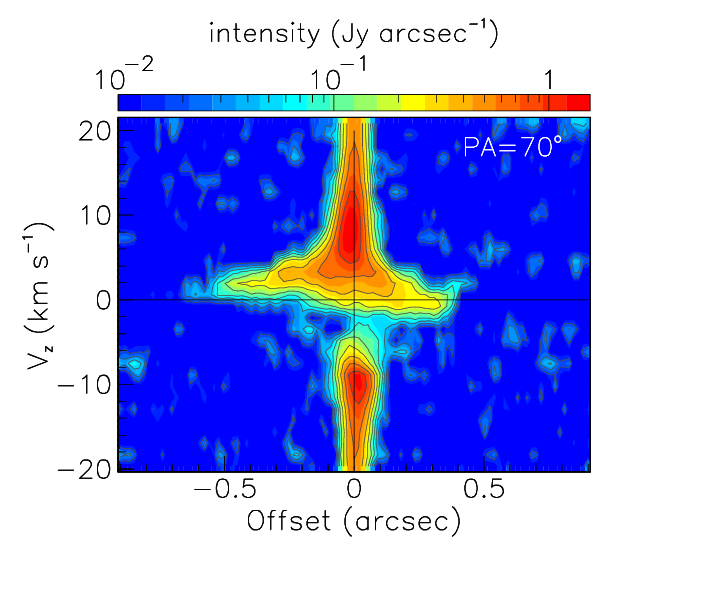}
  \includegraphics[width=4.8cm,trim=0.5cm .5cm 1.7cm 0cm,clip]{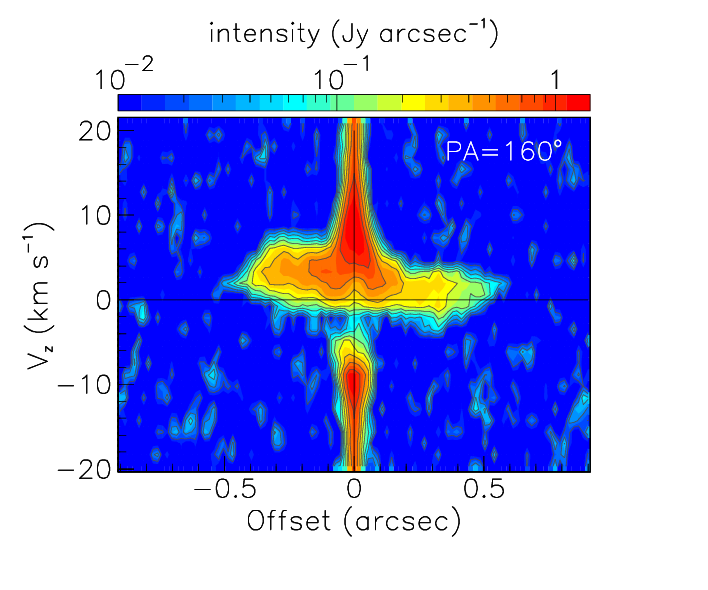}
  \includegraphics[width=4.3cm,trim=-.5cm -1.1cm 0cm 0cm,clip]{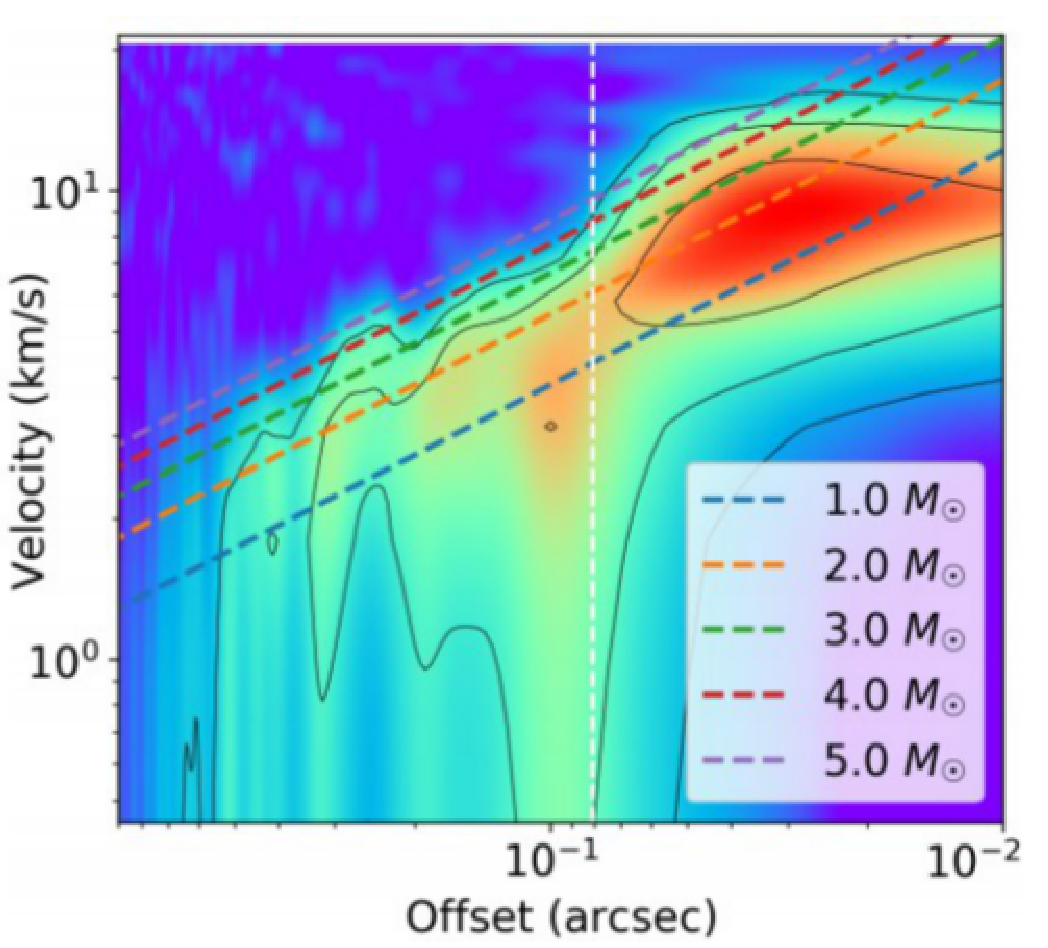}
  \caption{PV diagrams of the SiO(5-4) emission measured in 30 mas wide slits at position angles of 70\dego\ (left) and 160\dego\ (centre). The right panel \citep[Figure 14 of][reproduced with permission $\copyright$ ESO]{Homan2021} shows the red-eastern quadrant of the 70\dego\ diagram (left) in log-log scale. The dotted white line shows the distance from the star of 80 mas, at which \citet{Homan2021} locate the inner rim of a rotating disc. The other dotted lines illustrate the trends expected from Keplerian rotation associated with the presence of companions, the resulting central mass (star+companion) being indicated in the insert. }
 \label{fig8}
\end{figure*}

The interpretation of the same observations proposed by \citet{Homan2021} differs essentially by their interpreting the large Doppler velocities in the central region as the effect of Keplerian rotation rather than line broadening. These authors find that rotation dominates at projected distances $R$$\sim$80 mas, in agreement with our analysis. However, in contrast with our interpretation, they assume that this radius corresponds to the inner rim of a rotating disc that extends radially up to some 300 mas. Much of their argumentation is based on a PV diagram measured in a 30 mas wide slit oriented at a position angle of 70\dego. Figure \ref{fig8} displays this diagram together with that measured in a slit oriented perpendicularly, at position angle of 160\dego. At variance with \citet{Homan2021}, we have not subtracted the continuum contribution, being aware that the very strong opacity of the SiO line implies that accounting for the contribution of continuum emission cannot be done by a simple subtraction. Rotation is expected to contribute maximally to the 70\dego\ diagram and not to contribute to the 160\dego\ diagram. The comparison between the two diagrams illustrates well the difficulty to disentangle the respective contributions of rotation and line broadening. This is better seen in the right panel of the figure, copied from \citet{Homan2021}, which displays the red-eastern quadrant of the 70\dego diagram in log-log scale. At distances shorter than 80 mas, one sees clearly that the increase of the maximal $V_z$ with decreasing offset is the effect of line broadening described in Section \ref{sec3:ss2}: from $\sim$8 \kms\ to $\sim$18 \kms\ over the 80 to 30 mas range.  This is of course very different from the Keplerian trend and \citet{Homan2021} simply disregard this region in their interpretation. They concentrate instead on radial distances covering between $\sim$100 and $\sim$400 mas ($\sim$14 and $\sim$56 au) where whatever is left of the central rotation competes with the radial expansion of the inclined disc, as illustrated in Figure \ref{fig7}, where the first maximum of the $V_z$ distribution occurs near $\omega\sim$140\dego\ instead of $\sim$90\dego\ (our Figure \ref{fig4}) or $\sim$70\dego\ (for a rotation axis projecting at $\omega\sim-$20\dego) in the central region. Under such conditions it is difficult to reliably describe the rotation as Keplerian. This illustrates the importance of inspecting the whole data-cube over the region concerned, not simply a PV diagram measured over a narrow slit, in stars displaying both rotation and line broadening in the inner layers of their circumstellar envelope, such as L$_2$ Pup \citep{Kervella2016, Homan2017, Hoai2022} and R Dor \citep{Vlemmings2018, Homan2018, Nhung2021}. 

\begin{acknowledgements}

We thank Dr Ward Homan for useful discussions. This paper uses ALMA data ADS/JAO.ALMA\#2018.1.00659.L. ALMA is a partnership of ESO (representing its member states), NSF (USA) and NINS (Japan), together with NRC (Canada), MOST and ASIAA (Taiwan), and KASI (Republic of Korea), in cooperation with the Republic of Chile. The Joint ALMA Observatory is operated by ESO, AUI/NRAO and NAOJ. We are deeply indebted to the ALMA partnership, whose open access policy means invaluable support and encouragement for Vietnamese astrophysics. Financial support from the World Laboratory, the Odon Vallet Foundation and the Vietnam National Space Center is gratefully acknowledged. This research is funded by the Vietnam Academy of Science and Technology (VAST) under grant number NCVCC39.02/22-23.
\end{acknowledgements}

\label{lastpage}

\end{document}